\documentclass[12pt]{iopart}

\usepackage[T2A]{fontenc}
\usepackage[cp1251]{inputenc}
\usepackage{epsfig,graphicx}

\begin{document}

\title[Comparison of coherent Smith--Purcell radiation from metal and dielectric gratings]{Comparison of coherent Smith--Purcell radiation generated by 6.1 MeV electron beam in metal and dielectric lamellar gratings}

\author{L.G. Sukhikh, Yu.A. Popov, A.P. Potylitsyn}
\address{Applied Physics Department, Tomsk Polytechnic University, Lenin avenue, 2, Tomsk, 634050, Russia}
\ead{sukhikh@tpu.ru}

\author{G.A. Naumenko}
\address{Nuclear Physics Institute, Tomsk Polytechnic University, Lenin avenue, 2, Tomsk, 634050, Russia}

\begin{abstract}
Coherent Smith-Purcell radiation generated by bunched electron beam in the lamellar metal and dielectric gratings in the millimeter wavelength range was compared theoretically and experimentally. For theoretical estimation a simple model suitable for both dielectric and metal gratings was developed. Experimental comparison was carried out using extracted bunched 6.1 MeV electron beam of the microtron at Nuclear Physics Institute (Tomsk Polytechnic University). Both theoretical estimations and experimental data showed the difference of the radiation characteristics from the lamellar metal and dielectric gratings. The radiation from the dielectric grating had peak structure not monotonic one and was more intense comparing with metal grating radiation in the wavelength less than coherent threshold. These differences may be useful for research and development of new compact monochromatic radiation sources in sub-THz and THz region.

\end{abstract}

\maketitle

\section{Introduction}
Smith-Purcell radiation (SPR) arriving while a charged particle moves in the vicinity of the periodically deformed surface is widely used or planned to be used both for new Free Electron Laser (FEL) schemes~\cite{Walsh_FEL, Andrews0, Andrews, 5} including new terahertz sources~\cite{3,4,Prokop} and for beam  diagnostics~\cite{Doucas_NIMB_01,2,Doucas}. In these cases the metal gratings of the dif\mbox{}ferent shapes are used as radiators. For all metal gratings the SPR have a convenient feature that is so-called Smith-Purcell dispersion relation~\cite{1}:
\begin{equation}
\fl \displaystyle \lambda_m = \frac{d}{m} \Big (\beta^{-1} - \cos \theta \Big ), \quad m = 1, 2, \ldots
\label{Eq:1}
\end{equation}
Here $\lambda$ is the radiation wavelength, $d$ is the grating period, $m$ is the dif\mbox{}fraction order, $\beta =v/c$ is the particle velocity in the speed of light units, $\theta$ is the radiation polar angle. This relation was proven experimentally many times. In the case of coherent SPR from lamellar gratings the relation was demonstrated by Shibata et al. in~\cite{Shibata}.

Unfortunately, we still have a lack of knowledge about the properties of SPR from the dielectric gratings. The first theoretical investigations were made by Lampel in~\cite{Lampel} but we believe that the method used in cited paper is an ambiguous one.  Experimental investigations were made in the papers~\cite{Yamamoto_PC, Horiuchi} where authors examined the millimeter wavelength radiation from 2D and 1D photonic crystals used as SPR targets. The 1D photonic crystal is just a periodical structure, consisting of a number of teflon (PTFE) cylinders~\cite{Horiuchi}. Such structure is very similar to well-known grating and one may consider the radiation from it as SPR from the dielectric grating. Nevertheless, in the photonic crystals Cherenkov radiation (CR) plays an important role~\cite{Luo, Kremers}. 
The standard procedure to make theoretical studies of the radiation from the photonic crystal is to assume that the crystal is inf\mbox{}inite (a number of periods tends to infinity) and not absorbing one~\cite{Kremers}. That automatically gives us no chance to compare radiation characteristics from dielectric and metal gratings using the same formalism. In the millimeter wavelength region the grating have just some tens of periods and the inf\mbox{}inite assumption is not a good one. Hence, we need a model to simulate the characteristics of all kinds of radiation generated by the electron beam moving near the grating. Such model may be useful for updating of the radiation schemes for practical application.

In this paper we present the results of theoretical and experimental studies of the coherent SPR from the lamellar dielectric grating (or 1D photonic crystal) generated by the bunched electron beam with 6.1\,MeV energy in the millimeter wavelength region. Up to our knowledge that is only the second experimental investigation of such radiation, the f\mbox{}irst one was performed by Horiuchi et al. in~\cite{Horiuchi}. The main our goal is to develop a physically clear model suitable for a case of simultaneous generation of coherent SPR and CR in periodic structure with an arbitrary permittivity that makes it possible to compare the radiation characteristics from the dielectric and metal gratings. In our experimental part we  compare the characteristics of radiation from the aluminium and tef\mbox{}lon lamellar gratings. 
 
\section{Theoretical model}

In our theoretical estimations we will follow the recent papers of Karlovets and Potylitsyn~\cite{Karlovets_JETPLett_09, Karlovets_PRSTAB_10} where authors have shown a simple and elegant method of the Maxwell's equations solution that makes it possible to simulate the characteristics of any type of polarization radiation (including transition radiation, diffraction radiation, CR, SPR) appearing simultaneously. The term ``polarization radiation'' clearly shows a nature of the radiation that is a polarization of a target (grating) material by electromagnetic f\mbox{}ield of the traveling charged particle. In the cited articles~\cite{Karlovets_JETPLett_09, Karlovets_PRSTAB_10} it was shown that the method developed gives well known results for transition radiation, dif\mbox{}fraction one, CR and SPR that are just dif\mbox{}ferent ``kinematic'' cases of the polarization radiation.  

\begin{figure}[tb]
\centering
\includegraphics[width=85mm]{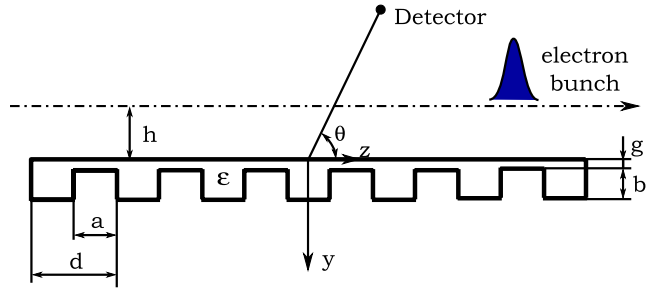}
\parbox{80mm}{\caption{Theoretical simulation scheme and some designations.\label{Teor_scheme}}}
\end{figure}

Fig.~\ref{Teor_scheme} shows the geometry of the problem and some designations. The electron bunch moves along $z$ axis with velocity $v$ and impact-parameter $h$. The grating is made of some material with permittivity  $\varepsilon(\omega)$ and have the period $d$ and the number of periods $N_d$. A groove width is $a$, depth is $b$, a substrate thickness is $g$. The grating size in direction perpendicular to the f\mbox{}igure plane is assumed to be inf\mbox{}inite. The detector is situated in the wave- (far-f\mbox{}ield) zone and its position is determined by a polar angle $\theta$ and an azimuth angle $\phi$. Such unusual grating orientation was chosen because we need to satisfy a condition that the radiation wavelength is less than the size of a surface of the grating through which we refract the radiation in order to use simlpe Fresnel coef\mbox{}f\mbox{}icients (see further). In cited paper~\cite{Karlovets_PRSTAB_10} this condition was satisfied in another way, it was assumed that $a \ll \lambda$.

According to the used method a magnetic f\mbox{}ield of the polarization radiation ${\bf H^{pol}} ({\bf r}, \omega )$ in a general case may be written as~\cite{Karlovets_JETPLett_09,Karlovets_PRSTAB_10}: 
\begin{eqnarray}
\fl \displaystyle {\bf H^{pol}} ({\bf r}, \omega ) = {\textrm {curl}} \frac{1}{c} \int \limits_{V_{T}} {\bf j_{pol}^{(0)}} ({\bf {r}}^{\prime}, \omega) 
\frac{e^{i \sqrt{\varepsilon (\omega )} \frac{\omega}{c} |{\bf r} - {\bf r}^{\prime}|}}{|{\bf r} - {\bf r}^{\prime}|} d^3 r^{\prime}. \label{4}
\end{eqnarray}
It should be mentioned that this formula is the exact solution of the Maxwell equation with the only assumption that the media is not magnetic one. Here  $c$ is the speed of light; ${\bf j_{pol}^{(0)}}({\bf {r}}^{\prime}, \omega) = \sigma (\omega) {\bf E_e} ({\bf {r}}^{\prime}, \omega)$ is the polarization current density, $\sigma (\omega) = \frac{(\varepsilon(\omega)-1) \omega}{4 \pi i}$ is the grating conductivity;  ${\bf E_e} ({\bf {r}}^{\prime}, \omega)$ is the Fourier transform of electron Coulomb f\mbox{}ield; $\frac{e^{i \sqrt{\varepsilon (\omega )} \omega |{\bf r} - {\bf r}^{\prime}|/c}}{|{\bf r} - {\bf r}^{\prime}|}$ is the Green function where ${\bf r}^{\prime}$ is the coordinate of the radiation point and ${\bf r}$ is the coordinate of the detection point. The integration is performed over the whole grating volume  $V_T$.

In our case (far-f\mbox{}ield radiation and infinite size in $x$ direction) the expression may be signif\mbox{}icantly simplif\mbox{}ied expanding Green function: 
\begin{eqnarray}
\begin{array}{l}
\fl\displaystyle  {\bf H^{pol}} ({\bf r}, \omega ) = \frac{2 \pi i}{c} \frac{e^{\sqrt{\varepsilon (\omega)}r \frac{\omega}{c}}}{r} {\bf k} \times \\ 
\fl \qquad \qquad \quad \times \displaystyle \int  dz^{\prime} dy^{\prime} {\bf j_{pol}^{(0)}} (k_x, y^{\prime}, z^{\prime}, \omega) e^{-i (k_y y^{\prime} + k_z z^{\prime})}.
\end{array}
 \label{15}
\end{eqnarray}
Here $\bf k$ is the wave-vector in the radiation direction, ${\bf j_{pol}^{(0)}} (k_x, y^{\prime}, z^{\prime}, \omega)$ is the special Fourier transform of the polarization current density. Physically such expansion means that we replace our radiating region inside the grating by a single ef\mbox{}fective dipole situated at the coordinate origin. 

Let us examine our bunched beam properties. Let us assume that our bunch have some longitudinal (along $z$ axis) distribution of $N_e$ ($N_e \gg 1$) noninteracting electrons that are moving in the same direction with the same speed. In this case a bunch current density may be written as:
\begin{equation}
\fl {\bf j} ({\bf r},t) = e {\bf v} \sum_{n=1}^{N_e} \delta (x)\,\delta (y-h)\, \delta (z-z_n-vt).
\end{equation}
Here $e$ is the elementary charge, ${\bf r_n} = \{0, h, z_n \}$ is the position of the  $n$-th electron in the bunch,  ${\bf v} = \{0,0,v \}$ is the bunch velocity vector. We do no take into account a transverse distribution of the electrons in the bunch because in the real experimental conditions (bunch transverse size is less than $\gamma \lambda$) a transverse form-factor is close to unity as it was shown by Shibata et al. in~\cite{Shibata}.

A complete Fourier transform of the bunch current density have the form:
\begin{equation}
\fl {\bf j} ({\bf k}, \omega) = \frac{e}{(2\pi)^3} \frac{{\bf v}}{v}\, e^{-i k_y h} \delta \left( \frac{\omega}{v} - k_z\right) \sum_n e^{-i k_z z_n}
\end{equation}

In this case the complete Fourier transform of the electron bunch f\mbox{}ield ${\bf E_e} ({\bf k},\omega)$ that is convenient to use obtaining a special Fourier transform ${\bf E_e} (k_x, y, z, \omega)$ that is needed for the problem solution (see Eq.~(\ref{15})) may be written as:
\begin{equation}
\fl {\bf E_e}({\bf k}, \omega) = \frac{2 e i}{(2 \pi)^2 \omega} \frac{{\bf v} \omega^2 /c^2- {\bf k} ({\bf k}, {\bf v} )}{k^2-\omega^2 / c^2}  e^{-i k_y h} \sum_n e^{-i k_z z_n }
\end{equation}
The special Fourier transform ${\bf E_e} (k_x, y, z, \omega)$ for a case $h<0$ may be found as:
\begin{eqnarray}
\displaystyle
\fl {\bf E_e}(k_x,y,z, \omega) = -\frac{i e}{2\pi v }  
\frac{\exp \left[-(h+ y)\sqrt{k_x^2 + \frac{\omega^2}{v^2} \gamma^{-2}} \right]} {\sqrt{k_x^2 + \frac{\omega^2}{v^2} \gamma^{-2}}} \times \\
\fl \times \left\{ k_x, -i \sqrt{k_x^2 + \frac{\omega^2}{v^2} \gamma^{-2}}, \frac{\omega}{v} \gamma^{-2} \right\} \sum_n  e^{i \frac{\omega} {v} (z-z_n)}.
\label{16}
\end{eqnarray}

Combining Eqs.~(\ref{15}) and~(\ref{16}) one may easily calculate double integral and obtain the radiation magnetic f\mbox{}ield:
\begin{eqnarray}
\begin{array}{l}
\displaystyle 
\fl {\bf H^{pol}} = -\frac{i}{2\pi c} \frac{\omega}{v}(\varepsilon -1)\frac{\exp \left[ i \sqrt{\varepsilon} \frac{\omega}{c} r \right]}{r} \frac{{\bf k} \times {\bf q}}{\sqrt{k_x^2 + \frac{\omega^2}{v^2}\gamma^{-2}}} \frac{\sin \left[ \frac{Nd}{2} \left(\frac{\omega}{v}-k_z \right) \right] }{\frac{\omega}{v}-k_z} 
\\
\fl \displaystyle  \left[ \frac{\sin \left[ \frac{d-a}{2} \left(\frac{\omega}{v}-k_z \right) \right] }{\sin \left[ \frac{d}{2} \left(\frac{\omega}{v}-k_z \right) \right]} \frac{\exp \left[-b \left(\sqrt{k^2_x+\frac{\omega^2}{v^2}\gamma^{-2}}+i k_y\right)\right] -1}{\sqrt{k^2_x+\frac{\omega^2}{v^2}\gamma^{-1}}+i k_y} + \right. \\
\fl \displaystyle 
\left. + \frac{\exp \left[-(g+b) \left(\sqrt{k^2_x+\frac{\omega^2}{v^2}\gamma^{-2}}+i k_y\right)\right] - \exp \left[-b \left(\sqrt{k^2_x+\frac{\omega^2}{v^2}\gamma^{-2}}+i k_y\right)\right]}{\sqrt{k^2_x+\frac{\omega^2}{v^2}\gamma^{-1}}+i k_y}  \right] \times \\
\fl \displaystyle \times \exp \left[ -h  \sqrt{k^2_x+\frac{\omega^2}{v^2}\gamma^{-2}} \right] \sum_n e^{-i \frac{\omega}{v} z_n}, \label{17}
\end{array}
\label{H_pol}
\end{eqnarray}
where we used a following designation:
\begin{eqnarray}
\fl \displaystyle {\bf q} = \left\{k_x, \sqrt{k_x^2+\frac{\omega^2}{v^2}\gamma^{-2}},\frac{\omega}{v}\gamma^{-2}    \right\}. \label{18}
\end{eqnarray}

A spectral-angular density of the radiation in vacuum (after refraction from the grating) from a single electron may be found like following:
\begin{equation}
\displaystyle
\fl \frac{d^2W_{s}}{\hbar d\omega d\Omega} = \frac{c r^2}{\hbar}|{\bf E^{R}_{vac}}|^2,
\label{dW}
\end{equation}
where in order to f\mbox{}ind the squared absolute value of the radiation electrical f\mbox{}ield $|{\bf E^{R}_{vac}}|^2$ we will use a reciprocity theorem~\cite{Landau} that was applied to the polarization radiation in~\cite{Karlovets_JETPLett_09,Karlovets_PRSTAB_10}.

\begin{equation}
\displaystyle
\fl |{\bf E^{R}_{vac}}|^2 = T_{\perp} |H^{pol}_{\perp}|^2 + T_{\parallel} \left( |H^{pol}_{\parallel}|^2 + |H^{pol}_{y}|^2  \right)
\label{E2}
\end{equation}
where $|H^{pol}_{\perp}|^2 $ and $|H^{pol}_{\parallel}|^2$ are the components of the magnetic f\mbox{}ield perpendicular and parallel to the incidence plane:
\begin{eqnarray}
\begin{array}{l}
\displaystyle
\fl H^{pol}_{\perp} = H^{pol}_x \frac{\sin \theta \sin \phi}{\sqrt{1-(\sin \theta \cos \phi)^2}} - H^{pol}_z \frac{\cos \theta}{\sqrt{1-(\sin \theta \cos \phi)^2}} \\
\displaystyle
\fl H^{pol}_{\parallel} = H^{pol}_x \frac{\cos \theta}{\sqrt{1-(\sin \theta \cos \phi)^2}} + H^{pol}_z \frac{\sin \theta \sin \phi}{\sqrt{1-(\sin \theta \cos \phi)^2}}
\end{array}
\label{H_comp}
\end{eqnarray}

\begin{eqnarray}
\begin{array}{l}
\displaystyle
\fl T_{\perp} = \left| \frac{2 \sin \theta \cos \phi}{\varepsilon \sin \theta \cos \phi + \sqrt{\varepsilon - 1 + (\sin \theta \cos \phi)^2}} \right|^2 \\
\displaystyle
\fl T_{\parallel} = \left| \frac{2 \sin \theta \cos \phi}{ \sqrt{\varepsilon} \left( \sin \theta \cos \phi + \sqrt{ \varepsilon - 1 + (\sin \theta \cos \phi)^2} \right) } \right|^2
\end{array}
\label{T_comp}
\end{eqnarray}
are the refraction coef\mbox{}f\mbox{}icients expressed through Fresnel ones. This expression is correct if the radiating surface is larger than the wavelength (see discussion in~\cite{Karlovets_PRSTAB_10}). In other way such assumption may give some errors. 

The components of the unit vector in radiation direction may be written taking into account Snell's law: 
\begin{equation}
\fl {\bf e} = \varepsilon^{-1/2}\left\{\sin \theta \, \sin \phi , -\sqrt{\varepsilon -1 +(\sin \theta \, \cos \phi)^2}  , \cos \theta \right\}
\label{unit_vec}
\end{equation}

Combining Eqs.~(\ref{H_pol}) --~(\ref{unit_vec}), one may obtain the solution of the problem. In the expression obtained there is, obviously, a squared sum of the radiation f\mbox{}ields of each electron that may be treated in the following way: 
\begin{eqnarray}
\displaystyle
\fl \left| \sum_{n=1}^{N_e} e^{-i \frac{\omega}{v} z_n} \right|^2 = \left\{
\begin{array}{l}
\displaystyle
N_e, \quad n=m; \\
\displaystyle
\sum_{n=1}^{N_e-1} e^{-i \frac{\omega}{v} z_n} \sum_{m=1}^{N_e} e^{i \frac{\omega}{v} z_m}, \quad n \neq m. 
\end{array}
\right.
\end{eqnarray}
In the case $n=m$ one obtains simple incoherent radiation that depends linearly on the bunch population. In the case $n \neq m$ one obtains coherent radiation:
\begin{eqnarray}
\begin{array}{l}
\fl  \displaystyle
\sum_{n=1}^{N_e-1} e^{-i \frac{\omega}{v} z_n} = \sum_{n=1}^{N_e-1} \int_{-\infty}^{\infty} \delta (z-z_n) e^{-i \frac{\omega}{v} z} dz =\\[3ex] 
\fl  \displaystyle
(N_e-1)\int_{-\infty}^{\infty} \rho (z) e^{-i \frac{\omega}{v} z} dz,
\end{array}
\end{eqnarray}
where for the Gaussian beam with rms $\sigma_z$ one may obtain ($N_e \gg 1$):
\begin{equation}
\fl  \rho(z)=\frac{1}{N_e-1} \sum_{n=1}^{N_e-1} \delta (z-z_n) \simeq \frac{1}{\sqrt{2\pi}\sigma_z} \exp\left[-\frac{z^2}{2 \sigma_z^2}\right]
\end{equation}

Total spectral-angular density of SPR radiation from the bunch with population $N_e$ may be written as:
\begin{equation}
\fl  \displaystyle
\frac{d^2W_{tot}}{\hbar d\omega\, d\Omega} = \frac{d^2W_{s}}{\hbar d\omega\, d\Omega} N_e \left[ 1+(N_e-1)  |f_z(\sigma_z)|^2 \right],
\label{33}
\end{equation}
where $\displaystyle \left|f_z(\sigma_z)\right|^2= \left|\int_{-\infty}^{\infty} \rho (z) e^{-i \frac{\omega}{v} z} dz\right|^2$ is a longitudinal form-factor of the electron bunch.
For the Gaussian bunch with rms $\sigma_z$ the form-factor have a standard form:
\begin{eqnarray}
\fl  \displaystyle
 |f_z(\sigma_z)|^2 = \exp \left[-\frac{\omega^2 \sigma_z^2}{\beta^2 c^2}\right].
\label{form}
\end{eqnarray}

During our theoretical estimations, we made several assumptions that should be taken into account during our experiment. First, we assumed that the width of the grating is inf\mbox{}inite. Second, we assumed that the point-liked detector is situated far away from the grating. Third important assumption is one concerning the radiation output through only one surface of the grating with single refraction, so we may loose some information about secondary refractions in the grooves.

Now let us to compare the characteristics of radiation from the teflon and metal gratings using the same expression~(\ref{33}). Let us assume that a teflon permittivity is $\varepsilon_t=2.1+0.001i$ and a metal permittivity is  $\varepsilon_m=1+10^6 i$. The period of both gratings is $d=12$\,mm, the number of periods is $N_d=13$, the groove width is $a=d/2=6$\,mm. For our teflon grating the groove depth is assumed to be $b_t=5$\,mm and substrate thickness is $g_t=1$\,mm. It is obvious that no SPR will be generated from the metal grating in the geometry shown in Fig.~\ref{Teor_scheme} because the polarization currents will be induced only in a skin-layer inside the substrate that have no periodic deformation. That is why we take the substrate thickness of the metal grating equal to $g_m=0$. This assumption may give us an error because of the previously used Fresnel coefficients but it is the only way to carry out the comparison using the same formalism. Because of the skin-layer we take the metal grating groove depth equal to $b_m=0.1$\,mm. The impact-parameter is equal $h=10$\,mm for approximately 6.1\,MeV ($\gamma=12$) electrons. The bunch length (rms) is $\sigma_z=2$\,mm. Both in simulations and in the experiment one component of radiation is taken into account, that is one in  $yz$ plane.

\begin{figure}[tb]
\centering
\includegraphics[width=80mm]{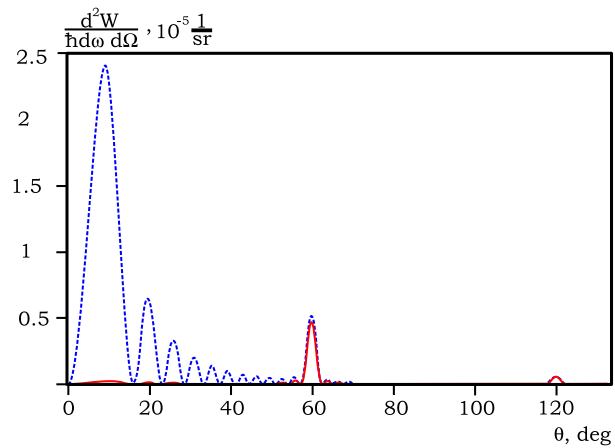}
\parbox{80mm}{\caption{Monochromatic ($\lambda=6$\,mm) coherent SPR from the teflon grating (blue dashed line) and the metal grating (red solid line).  \label{fig2}}}
\end{figure}

Fig.~\ref{fig2} shows the polar dependence of monochromatic ($\lambda=6$\,mm) coherent SPR from both gratings. One may clearly see in Fig.~\ref{fig2} the f\mbox{}irst and third orders of SPR line ($\theta_1=60$\,deg and $\theta_3=120$\,deg). The second order is suppressed and this moment is not clear. We believe this effect appears because of our single refraction assumption~\cite{Karlovets_PRSTAB_10}. Nevertheless, one may see that SPR from both gratings have almost the same intensity, but in the case of teflon grating there is a powerful radiation in the smaller polar angles. It is not really correct to speculate about forward directed radiation because during our theoretical simulations we do not take into account the one from the faces of the grating.

\begin{figure}[tb]
\centering
\includegraphics[width=80mm]{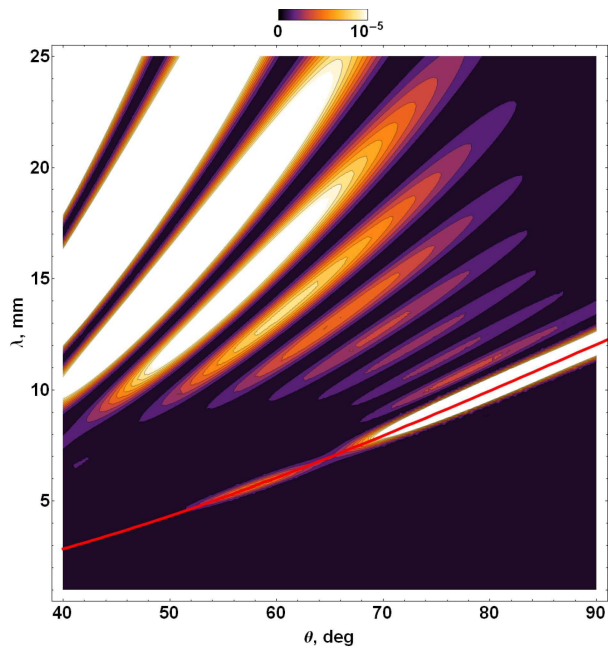}
\parbox{80mm}{\caption{Spectral--angular dependence of the coherent SPR from the teflon grating. Red line shows Smith--Purcell dispersion relation~(\ref{Eq:1}).  \label{fig3a}}}
\end{figure}

\begin{figure}[tb]
\centering
\includegraphics[width=80mm]{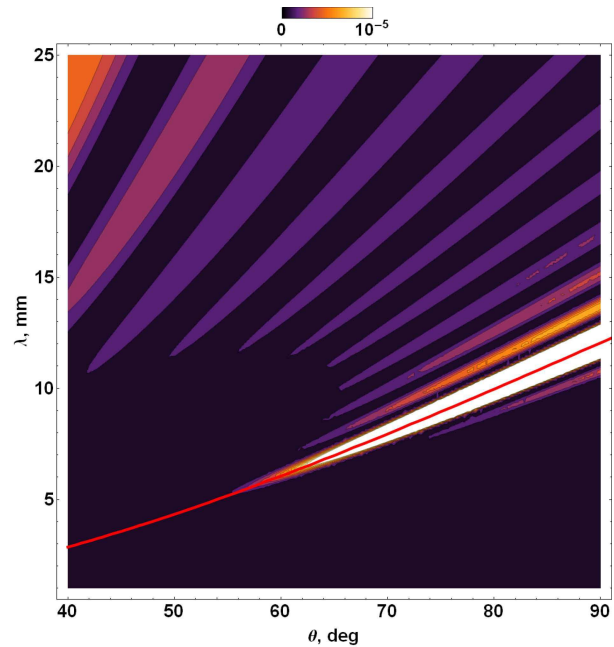}
\parbox{80mm}{\caption{Spectral--angular dependence of the coherent SPR from the metal grating. Red line shows Smith--Purcell dispersion relation~(\ref{Eq:1}).  \label{fig3b}}}
\end{figure}

Fig.~\ref{fig3a},~\ref{fig3b} show the spectral-angular distributions of the radiation from both gratings. One may clearly see that the red line, marking Smith--Purcell relation for the f\mbox{}irst dif\mbox{}fraction order is not the only dispersion relation in these f\mbox{}igures. The additional lines are very similar to the ones measured by Horiuchi et al. in~\cite{Horiuchi}. As it was mentioned before, the authors of the cited paper have measured SPR from the periodic teflon target and have found some additional dispersion lines that have not been found for the metal grating. The model developed shows almost the same situation: the additional radiation from the metal grating is very weak, comparing with teflon grating one. Such ef\mbox{}fect may be explained by the CR contribution. We also should mention that the simulated radiation is monotonic one. 

\section{Experiment}

\begin{figure}[tb]
\centering
\includegraphics[width=80mm]{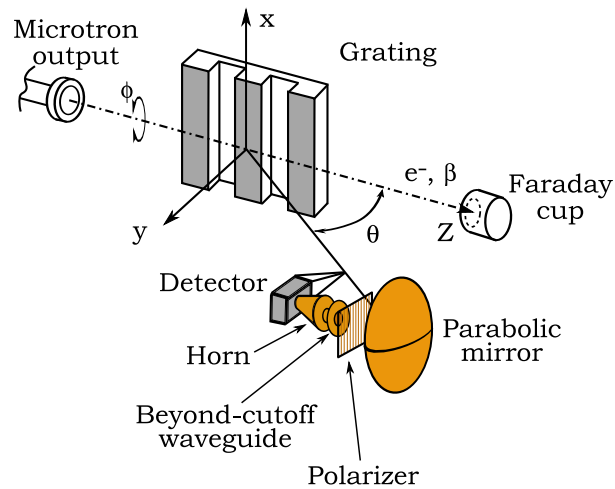}
\parbox{80mm}{\caption{The scheme of the experiment.  \label{fig:1_5}}}
\end{figure}

In order to check theoretical predictions we decided to carry out an experimental investigation of the spectral-angular characteristics of the coherent SPR from the teflon lamellar grating and to compare them with aluminium grating ones. The experimental scheme is shown in FIG.~\ref{fig:1_5}. The impact-parameter, polar angle $\theta$ and azimuth angle $\phi$ were changed during the experiment using stepper motors.

The electron beam extracted into air through $40$\,$\mu$m Be foil was used. The train of bunches with electron energy 6.1\,Mev ($\gamma\approx 12$), consisting of $n_b=10500$\,bunches (the bunch population is about $N_e=10^8$\,electrons) with $\tau=4$\,$\mu$s duration travels  in a line of the grating. The transverse sizes of the electron beam in the extraction point are about $4\times4$\,mm$^2$ (full width). The longitudinal distribution of the electron density in the bunch is believed to be a Gaussian with rms $\sigma_z=2$\,mm. 

The detecting system consisted of so-called ``telescope'', which represented a parabolic mirror (diameter $170$\,mm, focal distance $151$\,mm) in focus of which the detector was set up. Such telescope allows to measure the angular radiation characteristics equal to wave- (far-f\mbox{}ield) zone ones~\cite{Naum_JETPLett}. The distance from the grating to the parabolic mirror was equal to $450$\,mm.  The radiation from each train was detected using DP-21M detector. The last is based on wide-band antenna, high-frequency low barrier Schottky diode and preamplif\mbox{}ier. The sensitivity of the detector was measured in the wave regions $3.8 \div 5.6$\,mm and $11 \div 17$\,mm and was equal to $300$\,mV/mW at wavelengths $5.5$\,mm and $11.5$\,mm~\cite{17}. The registered waveband ($3 \div 25$\,mm) was limited by coherent threshold in the smaller wavelengths and by the beyond-cutof\mbox{}f waveguide (diameter $15$\,mm) used to decrease accelerator RF background in the larger wavelengths. An angular acceptance of the detector was def\mbox{}ined by the ratio of the beyond-cutof\mbox{}f waveguide diameter to the focal distance of the parabolic mirror, and was equal to about $5$\,deg. Incoherent radiation may not be measured by the detector.
The measured radiation yield was averaged over 20 trains. The statistical error was less than 10\% during the experiments. The beam center was def\mbox{}ined while scanning by the grating and measuring the Faraday cup signal. In this case the grating operated as a ``narrow scraper''. The latter was equal to $h=12$\,mm. During the experiment a grid polarizer was used and the radiation polarization component in yz plane was measured.

We used tef\mbox{}lon and aluminium lamellar gratings with the period $d=12$\,mm. Grating length was equal to $150$\,mm, width was equal to $120$\,mm, the groove width was $a=d/2=6$\,mm, the groove depth was $b=5$\,mm and substrate thickness was $g=1.3$\,mm. The scheme shown in Fig.~\ref{Teor_scheme} is not useful for the aluminium grating that is why it was set up in the ``standard'' position with grooves directed toward the detector. However, for the teflon grating both orientations are possible and they both were used during the experiment. Let us denote: G1 is the orientation of the teflon grating with the substrate directed toward the detector (just like in Fig.~\ref{Teor_scheme}) and G2 is the orientation of the teflon grating with the grooves directed toward the detector (opposite case).

\begin{figure}[tb]
\centering
\includegraphics[width=80mm]{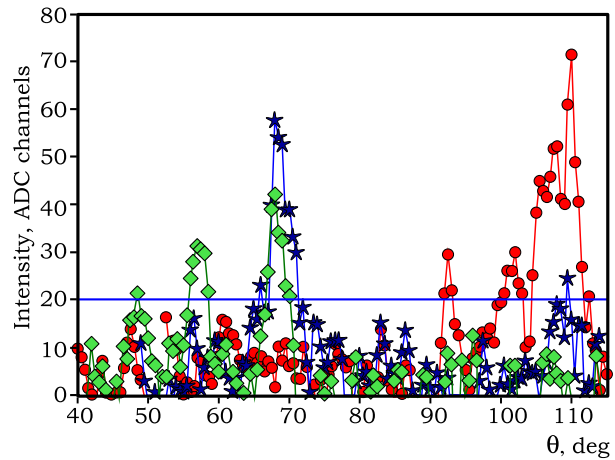}
\parbox{80mm}{\caption{Coherent SPR intensity vs. polar angle. Red dots -- aluminium grating, green diamonds -- teflon grating in $G1$ orientation, blue stars -- teflon grating in $G2$ orientation. Blue horizontal line shows the ``control level''.  \label{fig5}}}
\end{figure}

As the f\mbox{}irst step we measured the polar distribution of the coherent SPR for both gratings for all possible geometries. The result is shown in Fig.~\ref{fig5} by the red dots for aluminium grating, by the green diamonds for the teflon grating in G1 orientation and by the blue stars for the teflon grating in G2 orientation. The statistical error was comparable with the point size and is not shown in the f\mbox{}igure. Blue line shows a ``control level'' that was chosen in order to take into account only the stronger ef\mbox{}fects.

First one may see that almost all radiation from the aluminium grating is situated in the area of large polar angles and falls drastically while polar angle $\theta$ is larger than $112$\,deg. We believe that this fact that do not coincide with our theoretical estimations is caused by the f\mbox{}inite transverse size of the grating. Indeed, the transverse radius of the Lorentz-boosted electron f\mbox{}ield is about $\gamma \lambda$ and for $\lambda=16$\,mm (taken from Smith-Purcell relation for $\theta=110$\,deg) it is about $\gamma \lambda = 190$\,mm, while the grating size is just $120$\,mm. That seems to be a reason for signif\mbox{}icant attenuation of larger wavelengths. The radiation at polar angles less than $90$\,deg is very weak.

Let us analyze the SPR from the teflon grating. First of all one may see that the radiation is not monotonic one but have clear peak structure. We can see one good peak for G2 that is situated near $\theta=68$\,deg. For G1 we may see the same peak near $\theta=68$\,deg but it is not so intensive and we can see one additional peak near $\theta=57$\,deg. The peak structure looks very strange but it was also shown in the previously cited paper~\cite{Horiuchi}, where it was explained using the terms of photonic crystal band-gaps. In our theoretical estimations we assumed single refraction through one plane, but there seems to be an additional contribution to the radiation yield from the planes that were not taken into account (faces, grooves). It seems that this problem should be solved numerically. 

As the second step we measured the radiation spectra in both registered peaks. The low-pass dichoric f\mbox{}ilters were used for this procedure~\cite{Doucas,Hanke}. The f\mbox{}ilters were set up in order before the detector instead of beyond-cutof\mbox{}f waveguide. In this case the angular acceptance of the detector is rather large (about $16$\,deg) that plays a role in the detected spectra. 

\begin{figure}[tb]
\centering
\includegraphics{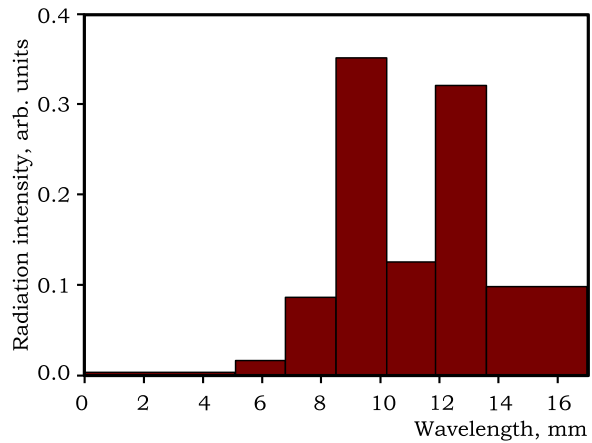}
\parbox{80mm}{\caption{The spectrum of the coherent SPR from the teflon grating with the orientation G2 in the peak near $68$\,deg. \label{fig6}}}
\end{figure}

In Fig.~\ref{fig6} the spectrum of the coherent SPR is shown. The spectrum was measured from the teflon grating with the orientation G2 in the peak near $68$\,deg. From Fig.~\ref{fig6} one may see that the measured spectrum have two lines: in a region from $8.5$ to $10.2$\,mm and from $11.9$ to $13.6$\,mm. According to the Smith-Purcell dispersion relation (Eq.~(\ref{Eq:1})) we should have here the wavelength $\lambda=7.5$\,mm. Such shift of the registered wavelength (f\mbox{}irst of them) is not really clear but the reason may be the large angular acceptance of the detector during the spectral measurements. The second spectral line has nothing to do with the Smith-Purcell relation. During our theoretical simulations (see Fig.~\ref{fig3a}) we shown that there are rather powerful additional radiation lines. It seems that the second measured spectral line comes from these additional lines.

\begin{figure}[tb]
\centering
\includegraphics{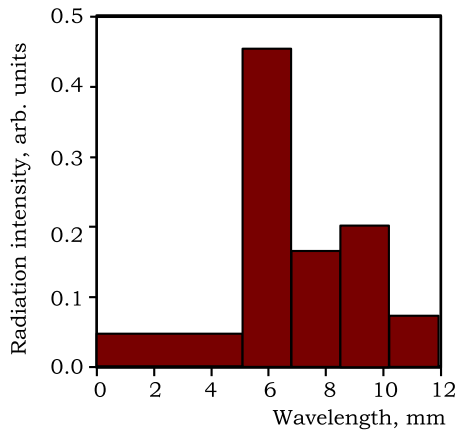}
\parbox{80mm}{\caption{The spectrum of the coherent SPR from the teflon grating with orientation G1 in the peak near $57$\,deg. \label{fig7}}}
\end{figure}

In Fig.~\ref{fig7} the spectrum of the coherent SPR that was measured from the teflon grating with orientation G1 in the peak near $57$\,deg is shown. This peak was measured only up to $12$\,mm wavelength because of the technical limitations. One may clearly see that the measured radiation have maximum in the wavelength region from $5.1$\,mm to $6.8$\,mm. According to the Smith-Purcell relation the radiation with wavelength $5.5$\,mm should be generated at this polar angle. This fact is the most interesting in  our investigation because this radiation wavelength is signif\mbox{}icantly lower than our coherent threshold. Substituting the wavelength ($\lambda=6$\,mm) and our bunch length ($\sigma_z=2$\,mm) into Eq.~(\ref{form}) one may obtain the value of form-factor that is equal to $|f_z(\sigma_z)|^2 = 0.012$. According to our theoretical simulations at this wavelength the coherent SPR from the metal grating and the teflon one should have almost same intensity (see Fig.~\ref{fig2}). But the radiation from the teflon grating is strong enough to be measured in a contrast to the radiation from the metal grating that was not detected (see Fig.~\ref{fig5}). This fact needs additional theoretical investigations but seems to be promising for creation of new radiation sources in sub-millimeter and terahertz regions based on coherent radiation of short electron bunches. One may also see in Fig.~\ref{fig7} that there is some amount of radiation in the region from $8.5$\,mm to $10.2$\,mm. The origin of this radiation is large angular acceptance of our detecting system: we detect some amount of radiation from 68\,deg peak.

\section{Conclusion}

In this paper we have presented the simple model for estimation of coherent SPR characteristics that is suitable for both metal and dielectric gratings. The predicted radiation from the dielectric grating dif\mbox{}fers drastically from the SPR from the metal one. There are additional intensive radiation lines because of CR mechanism. This additional radiation may be useful for the SPR based FELs because it may intensify the process of the continuous beam bunching. This assumption, surely, needs additional investigations.

In the experimental part of our investigation we have compared the characteristics of the coherent SPR from the aluminium and teflon lamellar gratings in the same conditions. This characteristics dif\mbox{}fer signif\mbox{}icantly: the radiation from the aluminium grating is situated in the area of large polar angles where it is expected because of the coherent threshold. The radiation from the teflon grating have clear peak structure, not monotonic one and these peaks are situated in the area of smaller polar angles. The intensity of radiation from the teflon grating and SPR intensity from the aluminum one have the same order of magnitude. The spectra of radiation in this peaks have complicated structure proving our estimation about additional radiation lines. Also the radiation spectrum in the peak $\theta=57$\,deg shows the presence of the $\lambda \sim 6$\,mm wavelength that should be significantly suppressed because of our coherent threshold. That means that the radiation from the teflon grating at this wavelength is strong enough to be measured in spite of the form-factor value equal to $0.012$ in contrast to the radiation from the metal grating that was not detected. This fact is also promising for creation of new radiation sources based on prebunched electron beams with small dimensions.

In the conclusion we may say that the polarization radiation from the dielectric gratings is a new f\mbox{}ield of investigations and seems to be really promising for practical application in various f\mbox{}ields.
 
\section{Acknowlegment}

The authors would like to thank D.V. Karlovets for useful discussion during theoretical estimations and the microtron staf\mbox{}f for the help during the experiment. The work was partly supported by the Russian Federal Agency for Education within the program ``Scientif\mbox{}ic and educational specialists of innovation Russia'' under the contracts No. $\Pi617$ and $\Pi1143$.

\end{document}